\documentclass[preprint,pra,showpacs,superscriptaddress]{revtex4-1}

\usepackage{physics,graphicx,amsmath,amssymb,amsthm,subfig,xcolor,tikz}

\begin{document}

\title{Geometry and symmetry in quantum Boltzmann machine}

\author{Hai-Jing Song}

\affiliation{Institute of Physics, Beijing National Laboratory for
  Condensed Matter Physics, Chinese Academy of Sciences, Beijing
  100190, China}

\affiliation{School of Physical Sciences, University of Chinese
  Academy of Sciences, Beijing 100049, China}

\author{Tieling Song}

\affiliation{Institute of Physics, Beijing National Laboratory for
  Condensed Matter Physics, Chinese Academy of Sciences, Beijing
  100190, China}

\author{Qi-Kai He}

\affiliation{Institute of Physics, Beijing National Laboratory for
  Condensed Matter Physics, Chinese Academy of Sciences, Beijing
  100190, China}

\affiliation{School of Physical Sciences, University of Chinese
  Academy of Sciences, Beijing 100049, China}

\author{Yang Liu}

\affiliation{QuantaEye Technologies Co., Ltd, Beijing 100083, China}

\author{D.~L. Zhou}

\email{zhoudl72@iphy.ac.cn}

\affiliation{Institute of Physics, Beijing National Laboratory for
  Condensed Matter Physics, Chinese Academy of Sciences, Beijing
  100190, China}

\affiliation{School of Physical Sciences, University of Chinese
  Academy of Sciences, Beijing 100049, China}

\begin{abstract}
  Quantum Boltzmann machine extends the classical Boltzmann machine
  learning to the quantum regime, which makes its power to simulate
  the quantum states beyond the classical probability distributions.
  We develop the BFGS algorithm to study the corresponding
  optimization problem in quantum Boltzmann machine, especially focus
  on the target states being a family of states with parameters. As an
  typical example, we study the target states being the real symmetric
  two-qubit pure states, and we find two obvious features shown in the
  numerical results on the minimal quantum relative entropy: First,
  the minimal quantum relative entropy in the first and the third
  quadrants is zero; Second, the minimal quantum relative entropy is
  symmetric with the axes $y=x$ and $y=-x$ even with one qubit hidden
  layer. Then we theoretically prove these two features from the
  geometric viewpoint and the symmetry analysis. Our studies show that
  the traditional physical tools can be used to help us to understand
  some interesting results from quantum Boltzmann machine learning.
\end{abstract}


\maketitle

\section{Introduction}
\label{sec:intro}

The central aim of machine learning is to implement a task by building
a mathematical model that simulates the learning processes of a human
being with a computer~\cite{Goldberg1988,Andrieu2003,Jordan255}. One
of the probabilistic machine learning models is the restricted
Boltzmann machine~\cite{ZHANG20181186,2017PhRvB..96t5152N}, which has
been successfully used as a deep machine learning model in diverse
tasks~\cite{18(7).1527-1554,2018arXiv180209558C,2162-237X}.

Recently in Ref.~\cite{PhysRevX.8.021050} M. H. Amin et al. proposed
the quantum version of Boltzmann machine with a transverse field Ising
Hamiltonian, and presented a training algorithm by sampling. In
particular, they discussed the possibility of the implementation of
such a Boltzmann machine in quantum annealing
processors~\cite{2015arXiv151006356A,PhysRevA.94.022308,2067} like
D-wave.

The main advantage of quantum Boltzmann machine is that it can
simulate a multipartite entangled quantum state, which is beyond the
classical probability distribution as shown in Bell's
inequality~\cite{2018arXiv180311278K,2015arXiv151202900A,2013arXiv1307.0411L}.
According to modern physics, nature is governed by quantum mechanics,
so it is necessary to extend the machine learning to the quantum
regime.

The central problem in training Boltzmann machine is to minimize the
quantum relative entropy~\cite{2018arXiv180401683S} for the target
state with respect to the states corresponding to the Boltzmann
distributions for a given type of Hamiltonians~\cite{500000443545}. We
solve this minimization problem by adopting the BFGS
algorithm~\cite{2012arXiv1212.5929Y,2017arXiv170306690G}, where a
recenter technique is adopted to increase the stability of our
algorithm.

It is worth pointing out that the problem of minimizing the quantum
relative entropy for the target state and the states of Boltzmann
distributions is also explored in  measuring
the correlations of an $n$-partite quantum state in
Refs.~\cite{PhysRevLett.101.180505,PhysRevA.80.022113}. Remarkably, the
algorithms of calculating the many-body correlation were proposed
in~\cite{1751-8121-46-12-125301,0253-6102-61-2-07,1367-2630-18-2-023024},
which are applicable in quantum Boltzmann machine without hidden layers.

As a typical application of the above BFGS algorithm, we calculate the
minimum quantum relative entropy for the target states being a family
of two-qubit states with two real parameters, a radius and an angle.
The minimal quantum relative entropy shows two obvious features:
strongly angle-dependent and symmetric with two axes.

In general, numerical results from machine learning are hard to
understand since the machine learning focuses on to dig out the
results but not to discover the underlying mechanism. Therefore it is
of great importance to grasp the underlying mechanism of the numerical
optimal results~\cite{ISBN0-387-98793-2}. To understand the above two
features in the numerical results, here we use a geometric method and
a symmetry anlalysis to analytically explains these features.

Here we emphasize the relation between the learning problem and the
optimization problem in quantum Boltzmann machine. In general, in a
learning problem, only the learning sample is used to optimization,
which is often a partial information for the target state; however, in
our optimization problem, we assume that the target state is
completely known, and our task is to adjust the parameters to get the
approximate state as good as possible. In other words, the
optimization problem gives the optimal capacity for the best
approximate state for the learning problem in quantum Boltzmann
machine~\cite{PhysRevA.96.062327,2015ConPh..56..172S}.

\section{Definition of quantum Boltzmann machine}
\label{sec:defin-quant-boltzm}

A quantum Boltzmann machine is composed by a bipartite quantum system,
where one subsystem is called visible, and the other is called
hidden~\cite{PhysRevX.8.021050}. The quantum state of the bipartite
system is a Boltzmann thermal equilibrium state for a specific type
Hamiltonian with variable parameters. The task of the quantum
Boltzmann machine is to make the quantum state of the visible
subsystem approximate a given target state as well as possible by
adjusting the parameters in the Hamiltonian.

We formulate the Boltzmann machine mathematically as follows. Let
$\mathcal{H}_{v}\otimes \mathcal{H}_{h}$ denote the product bipartite
Hilbert space. The target state $\sigma_{\ast}$ is defined on
$\mathcal{H}_{v}$. A quantum Boltzmann machine has a specific type of
Hamiltonians with parameters
\begin{equation}
  \label{eq:5}
  H(\mathbf{a}) = - \mathbf{a} \cdot \mathbf{O},
\end{equation}
where $\mathbf{a}$ is the vector of real parameters, and $\mathbf{O}$
is the vector of linearly independent Hermitian operators with zero
trace. The quantum state of the bipartite quantum system is the
Boltzmann thermal equilibrium state~\cite{1751-8121-46-12-125301}
\begin{equation}
  \label{eq:1}
  \rho(\mathbf{a}) = \frac{e^{- H(\mathbf{a})}}{\Tr(e^{- H(\mathbf{a})})} =
  \frac{e^{\mathbf{a} \cdot \mathbf{O}}}{\Tr(e^{\mathbf{a} \cdot
      \mathbf{O}})}.
\end{equation}
Then the reduced state of the visible subsystem is
\begin{equation}
  \label{eq:2}
  \sigma(\mathbf{a}) = \Tr_{h}(\rho(\mathbf{a})),
\end{equation}
where $\Tr_{h}$ denotes the trace over the hidden subsystem. The aim
of the quantum Boltzmann machine is to minimize
\begin{equation}
  \label{eq:102}
  S_{m}(\sigma_{\ast}) = \min_{\mathbf{a}} S(\sigma_{\ast}|\sigma(\mathbf{a})),
\end{equation}
where the quantum relative entropy is defined as
\begin{equation}
  \label{eq:4}
  S(\sigma|\sigma^{\prime}) = \Tr_{v}(\sigma(\ln \sigma - \ln \sigma^{\prime})).
\end{equation}
Here the quantum relative entropy is used as a measure of the degree
of approximation of one quantum state with
another~\cite{2017APS..MARA51005R,PhysRevLett.101.180505}.


\section{BFGS algorithm}
\label{sec:bfgs-algorithm}

In this section, we apply the Broyden-Fletcher-Goldfarb-Shanno (BFGS)
algorithm to solve the minimization problem specified by
Eq.~\eqref{eq:102}.

The basic idea of the BFGS algorithm is as follows. The BFGS algorithm
is iterative. First, choose an initial parameter vector
$\mathbf{a}^{0}$. In the $(k+1)$-th iteration, calculate
$\mathbf{a}^{k+1}$ from $\mathbf{a}^{k}$, which is realized in two
steps. In the first step, find out the direction from $\mathbf{a}^{k}$
to $\mathbf{a}^{k+1}$, which is determined by the second order Taylor
expansion of the quantum relative entropy at the point
$\mathbf{a}_{k}$:
\begin{equation}
  \label{eq:56}
  S_{q}(\sigma_{\ast}|\sigma(\mathbf{a})) =
  S(\sigma_{\ast}|\sigma(\mathbf{a}^{k}))  +    \mathbf{g}^{k \mathbf{T}}  \mathbf{d}^{k}  +
  \frac{1}{2}  \mathbf{d}^{k \mathbf{T}} H^{k} \mathbf{d}^{k},
\end{equation}
where the upper index $\mathbf{T}$ denotes the transpose operation,
and
\begin{align}
  \label{eq:8}
  \mathbf{d}^{k} & = \mathbf{a} - \mathbf{a}^{k}, \\
  \mathbf{g}^{k} & = \eval{\pdv{S(\sigma_{\ast}|\sigma(\mathbf{a}))}{\mathbf{a}}}_{\mathbf{a}^{k}}, \\
  H^{k} & =
          \eval{\pdv{S(\sigma_{\ast}|\sigma(\mathbf{a}_{k}))}{\mathbf{a}^{k}}{\mathbf{a}^{k
          \mathbf{T}}}}_{\mathbf{a}^{k}}.
\end{align}
If the Hessian matrix $H^{k}$ is symmetric and positive, then
$S_{q}(\sigma_{\ast}|\sigma(\mathbf{a}))$ in Eq.~\eqref{eq:56} has a minimum
where the parameter vector $\mathbf{a}^{k}_{q}$ satisfies
\begin{equation}
  \label{eq:9}
  \mathbf{d}^{k}_{q} = \mathbf{a}^{k}_{q} - \mathbf{a}^{k} = -  {(H^{k})}^{-1}  \mathbf{g}^{k}.
\end{equation}
In the second step, let
\begin{equation}
  \label{eq:121}
  \mathbf{a}^{k+1} - \mathbf{a}^{k} = - \lambda^{k} {(H^{k})}^{-1}  \mathbf{g}^{k}.
\end{equation}

Then use the line search algorithm to determine $\lambda^{k}$ by minimizing
\begin{equation}
  \label{eq:122}
  \min_{\lambda^{k}} S(\sigma_{\ast}|\sigma(\mathbf{a}^{k} - \lambda^{k} {(H^{k})}^{-1}  \mathbf{g}^{k})).
\end{equation}
The key element in the BFGS algorithm is to derive iteratively
$(H^{k})^{-1}$ from $\mathbf{d}^{k}$ and $\mathbf{g}^{k}$, which is
explained in Ref.~\cite{NW2006}.

As discussed above, it is helpful to get the analytical expression of
the first derivative of $S(\sigma_{\ast}|\sigma(\mathbf{a}))$. Let us start the
calculations with the case without the hidden subsystem. In this case,
the first order derivative is
\begin{align}
  \label{eq:10}
  \frac{\partial S(\sigma_{\ast}|\sigma(\mathbf{a}))}{\partial a_{i}}
  & = - \Tr \left( \sigma_{\ast} \frac{\partial}{\partial a_{i}} \ln
    \frac{e^{\mathbf{a} \cdot\mathbf{O}}}{\Tr \left( e^{{\mathbf{a} \cdot
    \mathbf{O}}} \right)} \right) \nonumber\\
  & = - \Tr \left( \sigma_{\ast} \frac{\partial}{\partial a_{i}}
    \left( \mathbf{a} \cdot \mathbf{O} - \Tr \left( e^{{\mathbf{a} \cdot
    \mathbf{O}}} \right) \right)  \right) \nonumber\\
  & = - \Tr(\sigma_{\ast} O_{i}) + \frac{\Tr \left( O_{i} e^{\mathbf{a} \cdot
    \mathbf{O}} \right)}{\Tr \left( e^{\mathbf{a} \cdot \mathbf{O}} \right)}
    \nonumber\\
  & = \Tr(\rho(\mathbf{a}) O_{i}) - \Tr(\sigma_{\ast} O_{i}).
\end{align}
The parameter vector $\mathbf{a}_{\ast}$ satisfies
\begin{equation}
  \label{eq:58}
  \Tr(\rho(\mathbf{a}) O_{i}) = \Tr(\sigma_{\ast} O_{i}).
\end{equation}
In other words, we only need to find the minimization of
Eq.~\eqref{eq:102} in the set of states given by Eq.~\eqref{eq:58}. In
fact, the parameter vector $\mathbf{a}_{\ast}$ can be determined by
\begin{equation}
  \label{eq:60}
  \max_{\mathbf{a}} S(\rho(\mathbf{a})),
\end{equation}
where $\mathbf{a}$ satisfies Eq.~\eqref{eq:58}, and the Von Neumann
entropy $S(\tau)=-\Tr(\tau\ln \tau)$. In this case
\begin{align}
  \label{eq:59}
  S(\sigma_{\ast}|\rho(\mathbf{a}_{\ast}))
  & = \Tr(\sigma_{\ast} \ln \sigma_{\ast}) - \Tr(\sigma_{\ast}
    \ln(\rho(\mathbf{a}_{\ast}))) \nonumber\\
  & = \Tr(\sigma_{\ast} \ln \sigma_{\ast}) - \Tr(\sigma_{\ast}
    (\mathbf{a}_{\ast} \cdot \mathbf{O} - \ln(\Tr(e^{\mathbf{a}_{\ast} \cdot
    \mathbf{O}})))) \nonumber\\
  & = \Tr(\sigma_{\ast} \ln \sigma_{\ast}) - \Tr(\rho(\mathbf{a}_{\ast})
    (\mathbf{a}_{\ast} \cdot \mathbf{O} - \ln(\Tr(e^{\mathbf{a}_{\ast} \cdot
    \mathbf{O}})))) \nonumber\\
  & = S(\rho(\mathbf{a}_{\ast})) - S(\sigma_{\ast}).
\end{align}

Now let us consider the Boltzmann machine with a hidden subsystem.
Then the first derivative of $S(\sigma_{\ast},\sigma(\mathbf{a}))$ is
\begin{align}
  \label{eq:7}
  \pdv{S(\sigma_{\ast}|\sigma(\mathbf{a}))}{a_{i}}
  & = - \pdv{a_{i}} \Tr_{v} \left(\sigma_{\ast} \ln
    \frac{\Tr_{h} \left(e^{ \mathbf{a}
    \cdot\mathbf{O}}\right)}{\Tr \left(e^{\mathbf{a} \cdot\mathbf{O}} \right)}\right)
    \nonumber\\
  & = -  \Tr_{v} \left(\sigma_{\ast} \pdv{a_{i}} \ln
    \frac{\Tr_{h} \left(e^{ \mathbf{a}
    \cdot\mathbf{O}}\right)}{\Tr \left(e^{\mathbf{a} \cdot\mathbf{O}} \right)}\right)
    \nonumber\\
  & = - \Tr_{v} \left(\sigma_{\ast} \pdv{a_{i}} \ln \Tr_{h} \left( e^{\mathbf{a}
    \cdot \mathbf{O}} \right)\right) + \pdv{a_{i}} \ln \Tr(e^{\mathbf{a} \cdot
    \mathbf{O}}) \nonumber\\
  & = \Tr(\rho(\mathbf{a}) O_{i}) - \Tr_{v} \left(\sigma_{\ast} \pdv{a_{i}} \ln \Tr_{h} \left( e^{\mathbf{a}
    \cdot \mathbf{O}} \right)\right).
\end{align}

In Section 3.3 of~\cite{Springer2014} Hiai Fumio and Petz D¨¦nes give
a convenient formula for the derivation (with respect to $ t\in R $):
\begin{equation}
  \frac{d}{dt} \log(A+tT) = \int_{0}^{\infty} dx (xI+A)^{-1}T(xI+A)^{-1} .
\end{equation}

Let
\begin{align}
  \label{eq:42}
  E(\mathbf{a}) & = e^{\mathbf{a} \cdot \mathbf{O}} = \sum_{l} e_{l}
                  |l\rangle\langle l|, \\
  D(\mathbf{a}) & = \Tr_{h} \left( E(\mathbf{a}) \right)
                  = \sum_{x} |x\rangle d_{x} \langle x|, \\
  B_{i}(\mathbf{a}) & = \pdv{D(\mathbf{a})}{a_{i}}.
\end{align}

Then we have
\begin{align}
  \label{eq:43}
  \pdv{S(\sigma_{\ast}|\rho(\mathbf{a}))}{a_{i}} & = \Tr(\rho(\mathbf{a}) O_{i}) - \int_{0}^{\infty} ds \Tr_{v}
                                          \left(\sigma_{\ast} \frac{1}{D + s} B_{i} \frac{1}{D + s} \right),
\end{align}
where the integral
\begin{align}
  \label{eq:47}
  & \phantom{=} \int_{0}^{\infty} ds \Tr_{v} \left(\sigma_{\ast}
    \frac{1}{D + s} B_{i} \frac{1}{D + s} \right) \nonumber\\
  & = \sum_{x,y} \int_{0}^{\infty} ds \frac{1}{d_{x} + s} \frac{1}{d_{y} +
    s} \Tr_{v} \left(\sigma_{\ast} |x\rangle\langle x| B_{i}
    |y\rangle\langle y| \right) \nonumber\\
  & = \sum_{x,y} \frac{\ln d_{y} - \ln d_{x}}{d_{y} - d_{x}}  \langle
    y| \sigma_{\ast} |x\rangle\langle x| B_{i}|y\rangle
\end{align}
with
\begin{align}
  \label{eq:79}
  B_{i}(\mathbf{a}) & = \pdv{D(\mathbf{a})}{a_{i}} \nonumber\\
                    & = \Tr_{h} \left(  \int_{0}^{1} d\tau
                      e^{ \tau \mathbf{a} \cdot\mathbf{O}} O_{i} e^{(1-\tau) \mathbf{a}
                      \cdot \mathbf{O}} \right) \nonumber\\
                    & = \sum_{l,m} \Tr_{h} \left( \int_{0}^{1} d\tau e_{l}^{\tau} e_{m}^{1 - \tau} |l\rangle\langle l| O_{i}
                      |m\rangle\langle m|\right) \nonumber\\
                    & = \sum_{l,m} \Tr_{h} \left( \frac{e_{l} - e_{m}}{\ln e_{l} - \ln e_{m}} |l\rangle\langle l| O_{i}
                      |m\rangle\langle m|\right).
\end{align}
Here we take the limit to deal with the cases of $d_{y}=d_{x}$ and
$e_{l}=e_{m}$:
\begin{align}
  \label{eq:49}
  \lim_{d_{y} \to d_{x}} \frac{\ln d_{y} - \ln d_{x}}{d_{y} - d_{x}} &
                                                                      =
                                                                      \frac{1}{d_{x}},
  \\
  \lim_{e_{m}\to e_{l}}  \frac{e_{l} - e_{m}}{\ln e_{l} - \ln e_{m}} &
                                                                      = e_{l}.
\end{align}
Since the above expression of the first order derivative of the
quantum relative entropy does not contain derivatives or integrals,
its numerical calculation becomes more effective and more accurate,
which significantly enhances the performance of the BFGS algorithm.

\section{Geometry in quantum Boltzmann machine without hidden
  subsystem}
\label{sec:geom-quant-boltzm}

Let us start from the quantum Boltzmann machine without the hidden
subsystem, and the visible subsystem composed by two qubits. The
Hamiltonian is given by~\cite{1751-8121-41-6-065005}
\begin{equation}
  \label{eq:71}
  H = -  \sum_{i=1}^{2} a_{i} X_{i} -  \sum_{i=1}^{2} a_{i+2} Z_{i}  -  a_{5} Z_{1} \otimes Z_{2},
\end{equation}
where $X$ and $Z$ are two Pauli matrices. The eigen problem of $Z$ is
given by $Z|n\rangle={(-1)}^{n}|n\rangle$ with $n\in\{0,1\}$, and the operator
$X$ is defined by $X|n\rangle=|1-n\rangle$.

To be concrete, the target states we study are the following set of
symmetric pure states
\begin{equation}
  \label{eq:83}
  |\psi(r,\phi)\rangle = \sqrt{1 - r^{2}} \; \frac{|01\rangle +
    |10\rangle}{\sqrt{2}} + r \cos \phi |00\rangle + r \sin \phi |11\rangle,
\end{equation}
where $r\in[0,1]$ and $\phi\in[0,2\pi]$.

\begin{figure}[htbp]
  \centering \includegraphics[width=8cm]{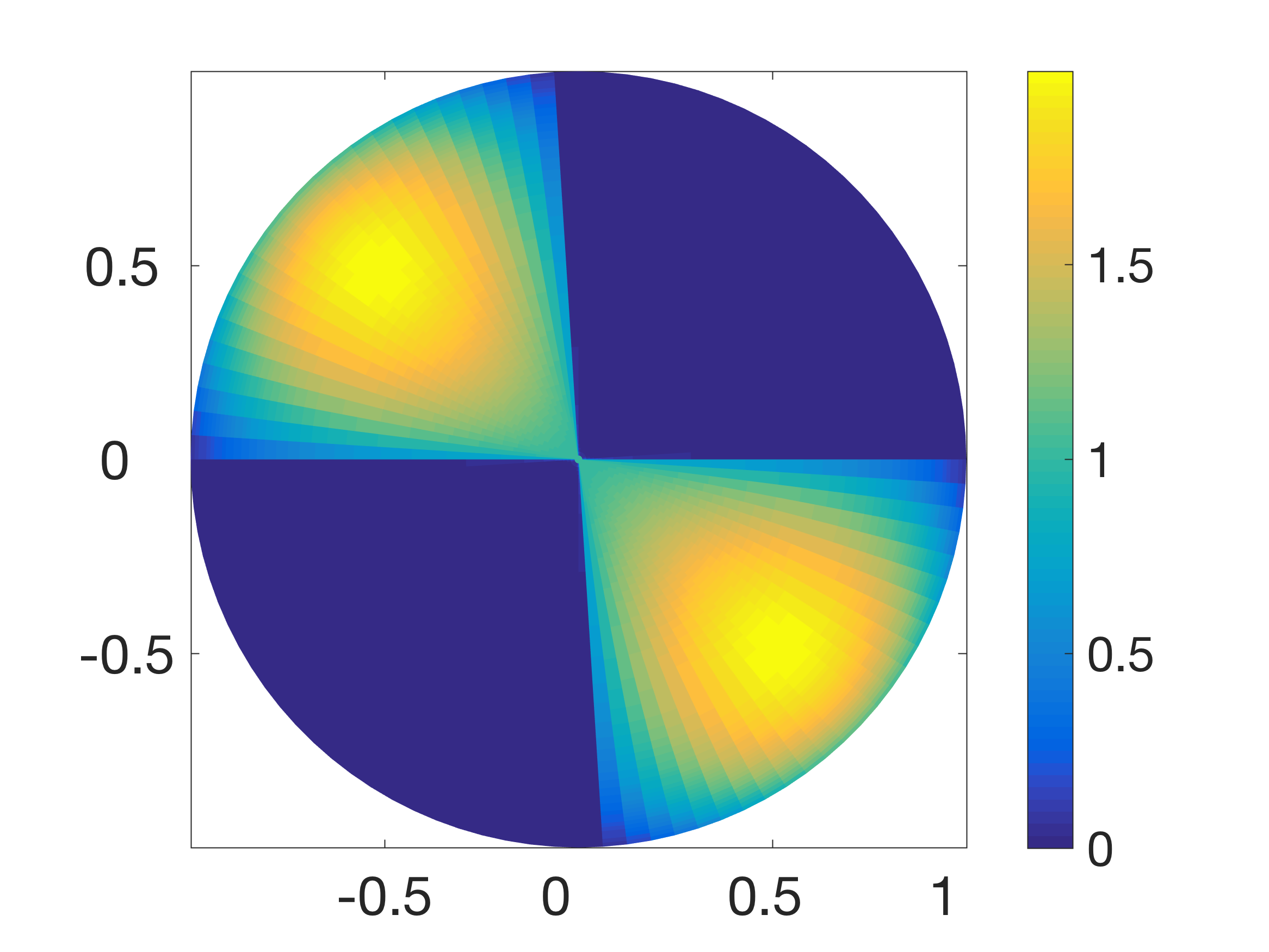}
  \caption{The minimum relative entropy for the states
    $|\psi(r,\phi)\rangle$. Here the coordinates $r,\phi$ are the polar coordinates.
    The minimum relative entropy is represented by the colors.}
  \label{fig:qbm2q0h}
\end{figure}

First, we numerically investigate the problem based on the BFGS
algorithm. The numerical result of the minimum quantum relative
entropy for the states $|\psi(r,\phi)\rangle$ is shown in Fig.~\ref{fig:qbm2q0h}.

We find the following four features of the numerical results from the
numerical results.
\begin{enumerate}
\item When the coordinates $(r,\phi)$ are in the first quadrant and in
  the third quadrant, the minimum relative entropy of the states
  $|\psi(r,\phi)\rangle$ to the exponential state family specified by
  $H$ are zero, i.e.
  \begin{equation}
    \label{eq:61}
    S_{m}(r,\phi) = 0, \text{if } 0<r<1 \text{ and }
    \phi\in(0,\frac{\pi}{2}) \text{ or } \phi\in(\pi,\frac{3\pi}{2}),
  \end{equation}
  which implies that every state in these two regions can be
  represented by one of the exponential states perfectly.

\item When $r=1$, the minimum quantum relative entropy $S_{m}(1,\phi)$ is
  periodic
  \begin{align}
    \label{eq:67}
    S_{m}(1,\phi+\frac{\pi}{2}) & =S_{m}(1,\phi), \\
    S_{m}(1,0) & = 0, \\
    S_{m}(1,\frac{\pi}{4}) & = 1.
  \end{align}

\item When $r=0$, the minimal quantum relative entropy
  $S_{m}(0, \phi)= 1$.

\item In addition, the maximal minimal quantum relative entropy is
  $2$, which appears at two states:
  \begin{equation}
    \label{eq:68}
    S_{m}\left(\frac{\sqrt{2}}{2}, \frac{3\pi}{4}\right)  =
    S_{m}\left(\frac{\sqrt{2}}{2}, \frac{7\pi}{4}\right) = 2.
  \end{equation}

\end{enumerate}

Why the minimum quantum relative entropy for the states
$|\psi(r,\phi)\rangle$ in the first quadrant and the third quadrant are zeros,
but they are nonzero for the states in the second and the fourth
quadrant?

To make the quantum relative entropy become zero, there must exist a
Hamiltonian $H$ in Eq.~\eqref{eq:71} such that
\begin{equation}
  \label{eq:87}
  |\psi(r,\phi)\rangle\langle\psi(r,\phi)| = \frac{e^{ H}}{\Tr(e^{H})}.
\end{equation}
The above equation seems to give the following paradox: the left side
is a pure state, while the right side is a mixed state, and the
equality can not be satisfied. This paradox can be solved as follows.
Since the state in the right hand side can be imagined as the thermal
equilibrium state of a Hamiltonian $H^{\prime}$ at temperature $T$ with
the relation
\begin{equation}
  \label{eq:88}
  H = - \frac{1}{k_{B} T} H^{\prime},
\end{equation}
where $k_{B}$ is the Boltzmann constant. When $H^{\prime}$ has a unique
ground state, the thermo-equilibrium state will limit to the ground
state at the zero temperature.

Thus, to prove the characteristic of the minimum quantum relative
entropy in Eq.~\eqref{eq:61}, we need to prove there exist a
$H^{\prime}$ with $|\psi(r,\phi)\rangle$ in the first and the third quadrant being
the unique ground state, and there does not exist such a $H^{\prime}$ for
$|\psi(r,\phi)\rangle$ in the second and the fourth quadrant.

For the quantum Boltzmann without hidden layers, we may show that
there is a unique $H$ such that the minimum quantum relative entropy
is arrived at. Since the target state $|\psi(r,\phi)\rangle$ is symmetric for the
permutation of qubit $1$ and qubit $2$, the corresponding Hamiltonian
must be symmetric with respect to the permutation. Without losing of
generality, we assume that the Hamiltonian
\begin{equation}
  \label{eq:89}
  H^{\prime} = a (Z_{1} + Z_{2}) + b (X_{1} + X_{2}) + Z_{1} \otimes Z_{2},
\end{equation}
namely, $a_{1}=a_{2}=b$, $a_{3}=a_{4}=a$, and $a_{5}=1$. The average
energy for the state $|\psi(r,\phi)\rangle$ is
\begin{align}
  \label{eq:90}
  E(a,b;r,\phi) & = \langle \psi(r,\phi)|H^{\prime}|\psi(r,\phi)\rangle
               \nonumber\\
             & = 2 a r^{2} \cos 2\phi + 4 b r \sqrt{1 - r^{2}}
               \sin(\phi + \frac{\pi}{4}) + 2 r^{2} - 1.
\end{align}
The necessary condition for $(r,\phi)$ being an extreme point is
\begin{align}
  \label{eq:91}
  \pdv{E(a,b;r,\phi)}{r} & = 4 a r \cos 2\phi + 4 b \sin(\phi +
                        \frac{\pi}{4}) \frac{1 - 2 r^{2}}{\sqrt{1 -
                        r^{2}}} + 4 r = 0, \\
  \pdv{E(a,b;r,\phi)}{\phi} & = - 4 a r^{2} \sin 2\phi + 4 b r \sqrt{1
                        - r^{2}} \cos(\phi + \frac{\pi}{4}) = 0.
\end{align}
Thus we get
\begin{align}
  \label{eq:92}
  a_{\ast} & = - \frac{(1 - r^{2}) \cot(\phi + \frac{\pi}{4})}{1 - 2 r^{2}
          \sin^{2} (\phi + \frac{\pi}{4})}, \\
  b_{\ast} & = - \frac{r \sqrt{1 - r^{2}} \sin 2\phi \csc(\phi +
          \frac{\pi}{4})}{1 - 2 r^{2} \sin^{2} (\phi + \frac{\pi}{4})}.
\end{align}
The Hessian matrix~\cite{10.1093/comjnl/7.4.308}
\begin{align}
  \label{eq:93}
  H & =
      \begin{pmatrix}
        \pdv[2]{E(a,b;r,\phi)}{r} &     \pdv{E(a,b;r,\phi)}{r}{\phi} \\
        \pdv{E(a,b;r,\phi)}{\phi}{r} & \pdv[2]{E(a,b;r,\phi)}{r}
      \end{pmatrix} \nonumber\\
    & =
      \begin{pmatrix}
        4 a \cos 2 \phi + \frac{4 b r \sin(\phi + \frac{\pi}{4}) (2 r^{2}
          -3)}{(1 - r^{2})^{3/2}} + 4 & - 8 a r \sin 2 \phi + 4 b \cos(\phi
        + \frac{\pi}{4}) \frac{1 - 2
          r^{2}}{\sqrt{1 - r^{2}}} \\
        - 8 a r \sin 2 \phi + 4 b \cos(\phi + \frac{\pi}{4}) \frac{1 - 2
          r^{2}}{\sqrt{1 - r^{2}}} & - 8 a r^{2} \cos 2\phi - 4 b r
        \sqrt{1 - r^{2}} \sin(\phi + \frac{\pi}{4})
      \end{pmatrix}
\end{align}
Then the Hessian matrix at the extreme point $(a_{\ast},b_{\ast})$
\begin{equation}
  \label{eq:94}
  H_{a_{\ast},b_{\ast}} =
  \begin{pmatrix}
    \frac{4 \sin 2 \phi}{(1 - r^{2}) (1 - r^{2} (1 + \sin 2\phi))} & \frac{4
      r \sin 2 \phi \cot(\phi + \frac{\pi}{4})}{1 - r^{2} (1 + \sin
      2\phi)} \\
    \frac{4 r \sin 2 \phi \cot(\phi + \frac{\pi}{4})}{1 - r^{2} (1 + \sin
      2\phi)} & \frac{4 r^{2} (1 - r^{2}) (2 - \sin 2\phi)}{1 - r^{2} (1 +
      \sin 2\phi)}
  \end{pmatrix}
\end{equation}
The determinant
\begin{equation}
  \label{eq:95}
  \mathrm{Det}(H_{a_{\ast},b_{\ast}}) = \frac{32 r^{2} \sin 2 \phi}{(1 + \sin 2
    \phi) (1 - r^{2} (1 + \sin 2 \phi))^{2}}.
\end{equation}
When $\mathrm{Det}(H_{a_{\ast},b_{\ast}})>0$, i.e.
$\phi\in(0,\frac{\pi}{2})$ or $\phi\in(\pi,\frac{3\pi}{2})$, the Hamiltonian
$H_{a_{\ast},b_{\ast}}$ is positive or negative, and
$E(a_{\ast},b_{\ast}; r,\phi)$ is the local minimum or maximum. In addition,
when $\phi\in(\frac{\pi}{2},\pi)$ or $\phi\in(\frac{3\pi}{2},2\pi)$,
$\mathrm{Det}(H_{a_{\ast},b_{\ast}})<0$, which implies that
$E(a_{\ast},b_{\ast}; r,\phi)$ is a saddle point. Thus for
$|\psi(r,\phi)\rangle$ in the first and the third quadrant there exists a
Hamiltonian $H^{\prime}$ (or $-H^{\prime}$) in Eq.~\eqref{eq:89} with the
state as its unique ground state. However, for $|\psi(r,\phi)\rangle$ in the
second and the fourth quadrant there does not exist such a Hamiltonian
$H$.

\begin{figure}[htbp]
  \centering \subfloat[][]{\includegraphics[width=6cm]{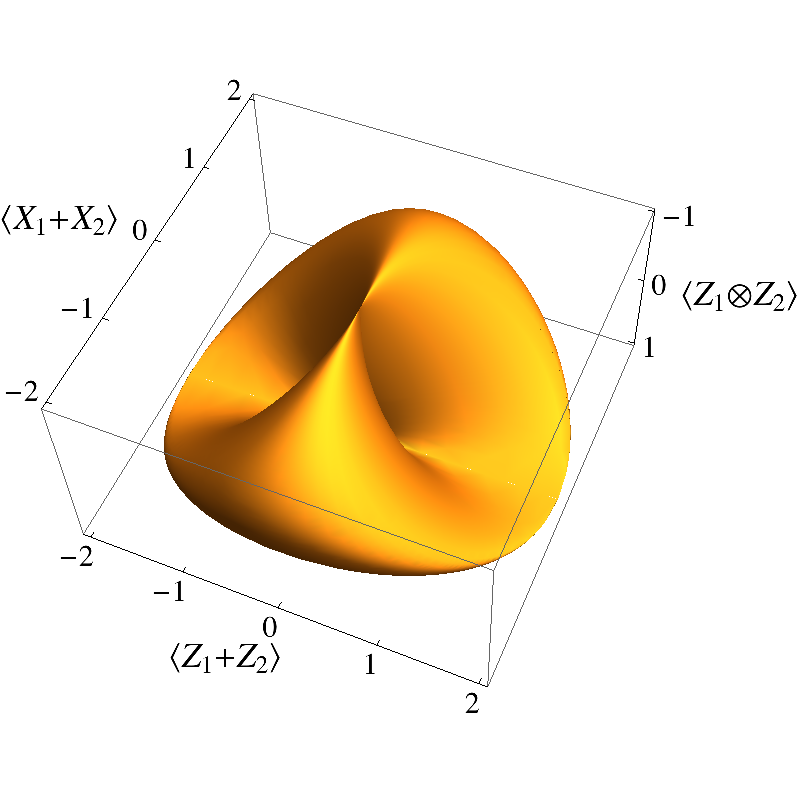}} \quad
  \subfloat[][]{\includegraphics[width=6cm]{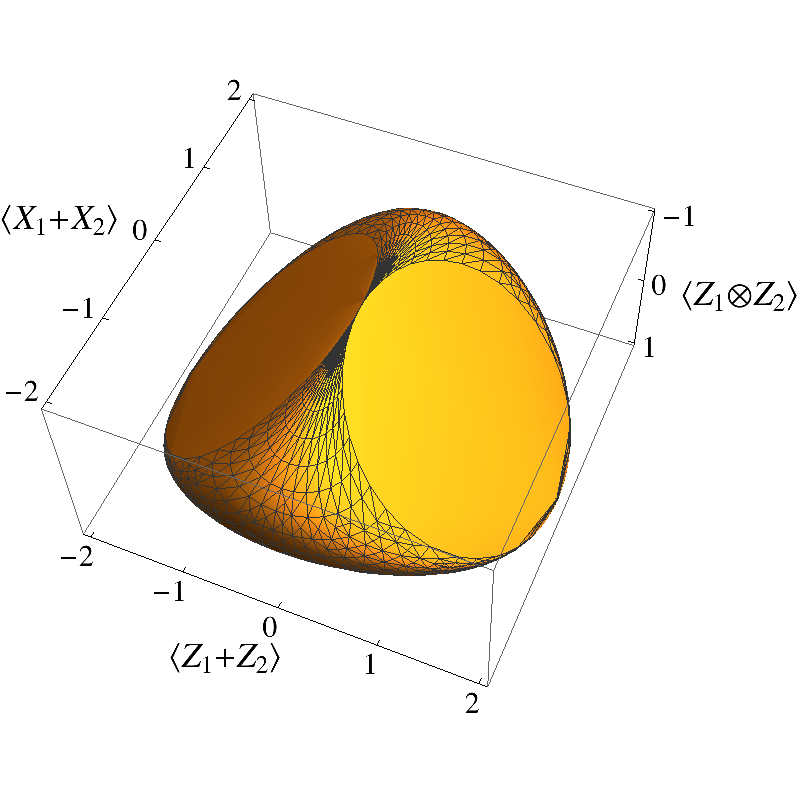}}
  \caption{The numerical range of the Hermitian operators
    $\{Z_{1}+Z_{2}, X_{1}+X_{2}, Z_{1}\otimes Z_{2}\}$ and the black net
    lines corresponding to the states $|\psi(r,\phi)\rangle$ in the first and the
    third quadrants.~\label{fig:qbm2q0hnr}}
\end{figure}

The above analysis can be clearly demonstrated in the geometric
pictures shown in
Fig.~\ref{fig:qbm2q0hnr}~\cite{2014arXiv1412.7642Z,Chen2016,1367-2630-17-8-083019}.
The average values of the Hermitian operators
$\{Z_{1}+Z_{2},X_{1}+X_{2},Z_{1}\otimes Z_{2}\}$ for the states
$|\psi(r,\phi)\rangle$ are plotted as a closed surface in
Fig.~\ref{fig:qbm2q0hnr}~(a), and the average values of the same three
operators are plotted as a convex body shown in
Fig.~\ref{fig:qbm2q0hnr}~(b). In particular, for the states
$|\psi(r,\phi)\rangle$ in the first and the third quadrant, the corresponding
points lie in the surface of the numerical range, as shown by the
black network in Fig.~\ref{fig:qbm2q0hnr}~(b). For the states
$|\psi(r,\phi)\rangle$ in the second and the fourth quadrant, the corresponding
points enter the interior of the convex body.

Note that the Hamiltonian in Eq.~\eqref{eq:89} and a constant $e$
defines a hyperplane specified by
\begin{equation}
  \label{eq:6}
  a \langle Z_{1} + Z_{2} \rangle + b \langle X_{1} + X_{2} \rangle +
  \langle Z_{1} \otimes Z_{2} \rangle = e.
\end{equation}
At any point corresponding to the state $|\psi(r,\phi)\rangle$ in the first or
the third quadrant, there is a unique hyperplane specified by
constants $a,b$ and $e$ tangent with the convex
body~\cite{Schneider1982}. The constants $a,b$ define the Hamiltonian
$H^{\prime}$ or $-H^{\prime}$ with $|\psi(r,\phi)\rangle$ as its unique ground state.
Hence we show that the minimum quantum relative entropy is zero for
the state $|\psi(r,\phi)\rangle$ in the first and third quadrant. For the state
$|\psi(r,\phi)\rangle$ in the second and fourth quadrant lying in the interior of
the convex body, there is no such a tangent hyperplane, and thus the
minimal quantum relative entropy becomes nonzero.

\begin{figure}[ht]
  \centering \subfloat[][]{\includegraphics[width=6cm]{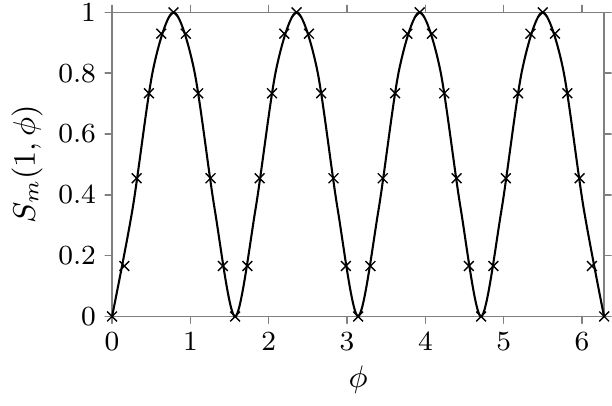}}
  \quad \subfloat[][]{\includegraphics[width=6cm]{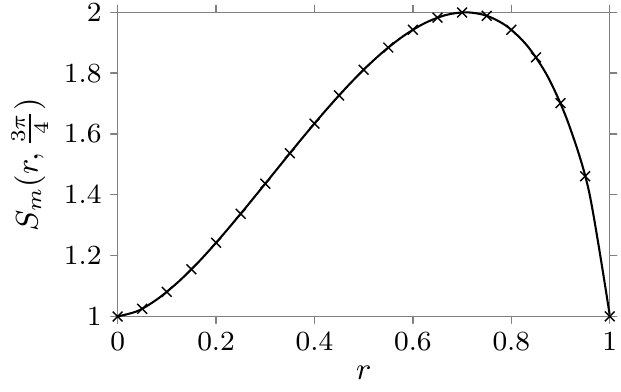}}

  \caption{The minimal quantum relative entropy in the two cases: (a)
    $r=1$; (b) $\phi=\frac{3\pi}{4}$. The analytical results are plotted
    as the solid lines, and the numerical results are plotted as cross
    marks. \label{fig:ex2}}
\end{figure}

Let us consider the states $|\psi(1,\phi)\rangle$. The average values
\begin{align}
  \label{eq:11}
  \ev{X_{1} + X_{2}}{\psi(1,\phi)} & = 0, \\
  \ev{Z_{1} + Z_{2}}{\psi(1,\phi)} & = 2 \cos(2\phi), \\
  \ev{Z_{1}\otimes Z_{2}}{\psi(1,\phi)} & = 1.
\end{align}
Then all the two-qubit states that satisfy the above conditions can be
written as
\begin{equation}
  \label{eq:12}
  \tau = \frac{I}{4} + \cos(2\phi) \frac{Z_{1}+Z_{2}}{4} +
  \frac{Z_{1}\otimes Z_{2}}{4} +  \ldots
\end{equation}
In terms of the basis specified the common eigenstates of $Z_{1}$ and
$Z_{2}$, the diagonal state
\begin{equation}
  \label{eq:17}
  \mathrm{diag}(\tau) = \frac{I}{4} + \cos(2\phi) \frac{Z_{1}+Z_{2}}{4} +
  \frac{Z_{1}\otimes Z_{2}}{4}.
\end{equation}
Because the Von Neaumann entropy for any quantum state is not more
than that of its diagonal state. According to Theorem 11.9 of Section 11.3.3 in~\cite{Nielsen:2011:QCQ:1972505},we have
\begin{equation}
  \label{eq:18}
  S(\tau) \le S(\mathrm{diag}(\tau)).
\end{equation}
Thus the minimal quantum relative entropy is
\begin{equation}
  \label{eq:19}
  S_{m}(1,\phi) = S(\mathrm{diag}(\tau)) = - \cos^{2}\phi \ln (\cos^{2} \phi) -
  \sin^{2}\phi \ln(\sin^{2} \phi).
\end{equation}
The analytical result in Eq.~\eqref{eq:19} is demonstrated in
Fig.~\ref{fig:ex2}~(a), together with the numerical result. These two
results from the BFGS algorithm agree well with each other.

Now we consider the states $|\psi(r,\frac{3\pi}{4})\rangle$. The average values
\begin{align}
  \label{eq:20}
  \ev{X_{1} + X_{2}}{\psi(r,\frac{3\pi}{4})} & = 0, \\
  \ev{Z_{1} + Z_{2}}{\psi(1,\frac{3\pi}{4})} & = 0, \\
  \ev{Z_{1}\otimes Z_{2}}{\psi(1,\frac{3\pi}{4})} & = 2 r^{2} - 1.
\end{align}
Then all the two-qubit states that satisfy the above conditions can be
written as
\begin{equation}
  \label{eq:21}
  \tau = \frac{I}{4} + (2 r^{2} - 1)
  \frac{Z_{1}\otimes Z_{2}}{4} +  \ldots
\end{equation}
In terms of the basis specified the common eigenstates of $Z_{1}$ and
$Z_{2}$, the diagonal state
\begin{equation}
  \label{eq:22}
  \mathrm{diag}(\tau) = \frac{I}{4} + (2 r^{2} - 1)
  \frac{Z_{1}\otimes Z_{2}}{4}.
\end{equation}
Because the Von Neaumann entropy for any quantum state is not more
than that of its diagonal state. According to Theorem 11.9 of Section
11.3.3 in~\cite{Nielsen:2011:QCQ:1972505}, we have
\begin{equation}
  \label{eq:23}
  S(\tau) \le S(\mathrm{diag}(\tau)).
\end{equation}
Thus the minimal quantum relative entropy is
\begin{equation}
  \label{eq:24}
  S_{m}(r,\frac{3\pi}{4}) = S(\mathrm{diag}(\tau)) = - r^{2} \ln r^{2} - (1
  - r^{2}) \ln (1 - r^{2}) + 1,
\end{equation}
which is demonstrated in Fig.~\ref{fig:ex2}~(b) supported by the
numerical results from the BFGS algorithm.

\section{Symmetry in quantum Boltzman machine}
\label{sec:symm-quant-boltzm}

In this section, we will explore to investigate the power of the
hidden subsystem in improving the capacity of the quantum Boltzmann
machine. More precisely, we study a three-qubit system with the third
qubit being the hidden subsystem, whose Hamiltonian is
\begin{equation}
  \label{eq:84}
  H = - \sum_{i=1}^{3} a_{i} X_{i} - \sum_{i=1}^{3} a_{i+3} Z_{i} -
  a_{7} Z_{2} \otimes Z_{3} - a_{8} Z_{1} \otimes Z_{3} - a_{9} Z_{1}
  \otimes Z_{2}.
\end{equation}

\begin{figure}[htbp]
  \subfloat[][]{\includegraphics[width=6cm]{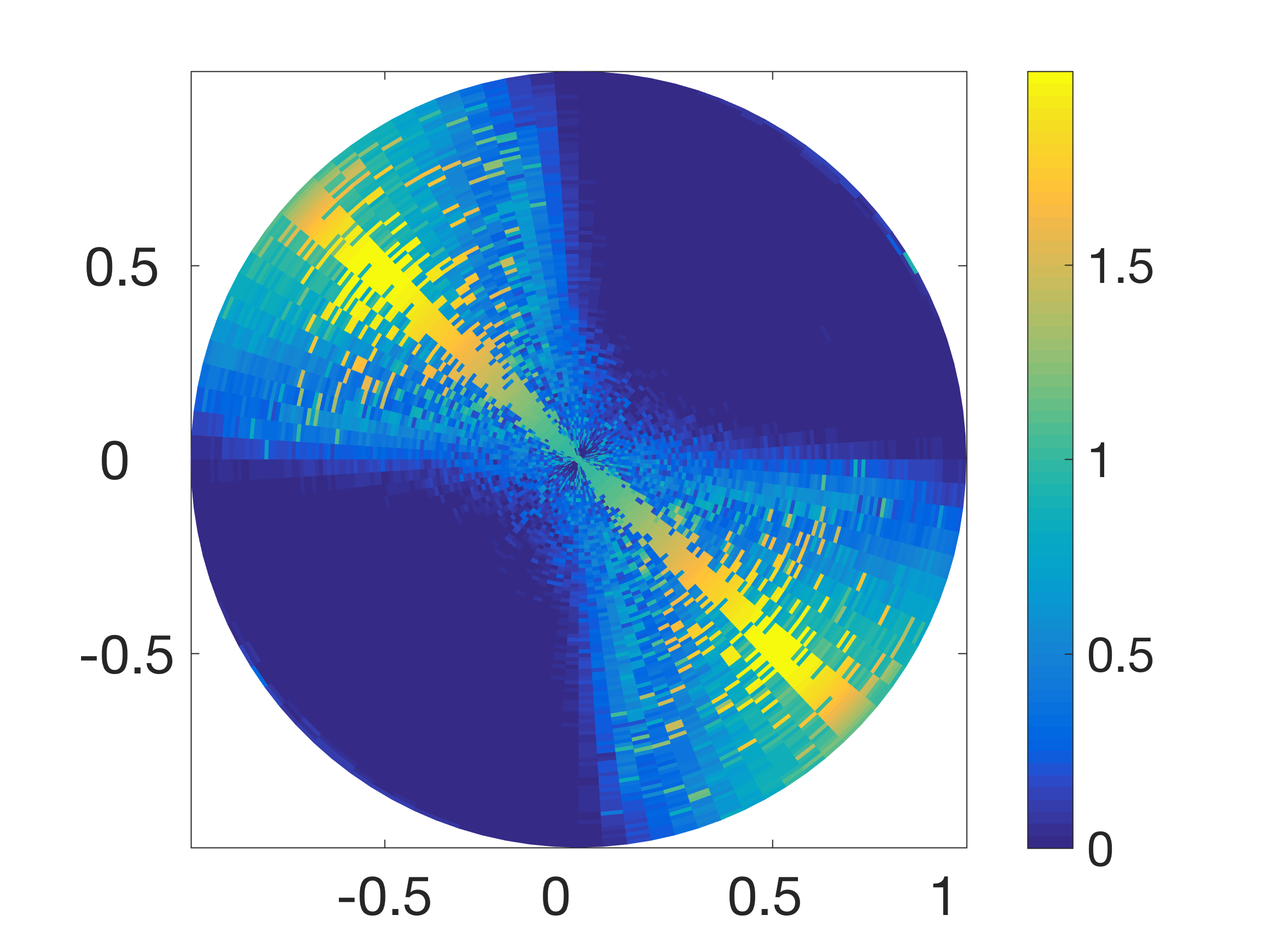}} \quad
  \subfloat[][]{\includegraphics[width=6cm]{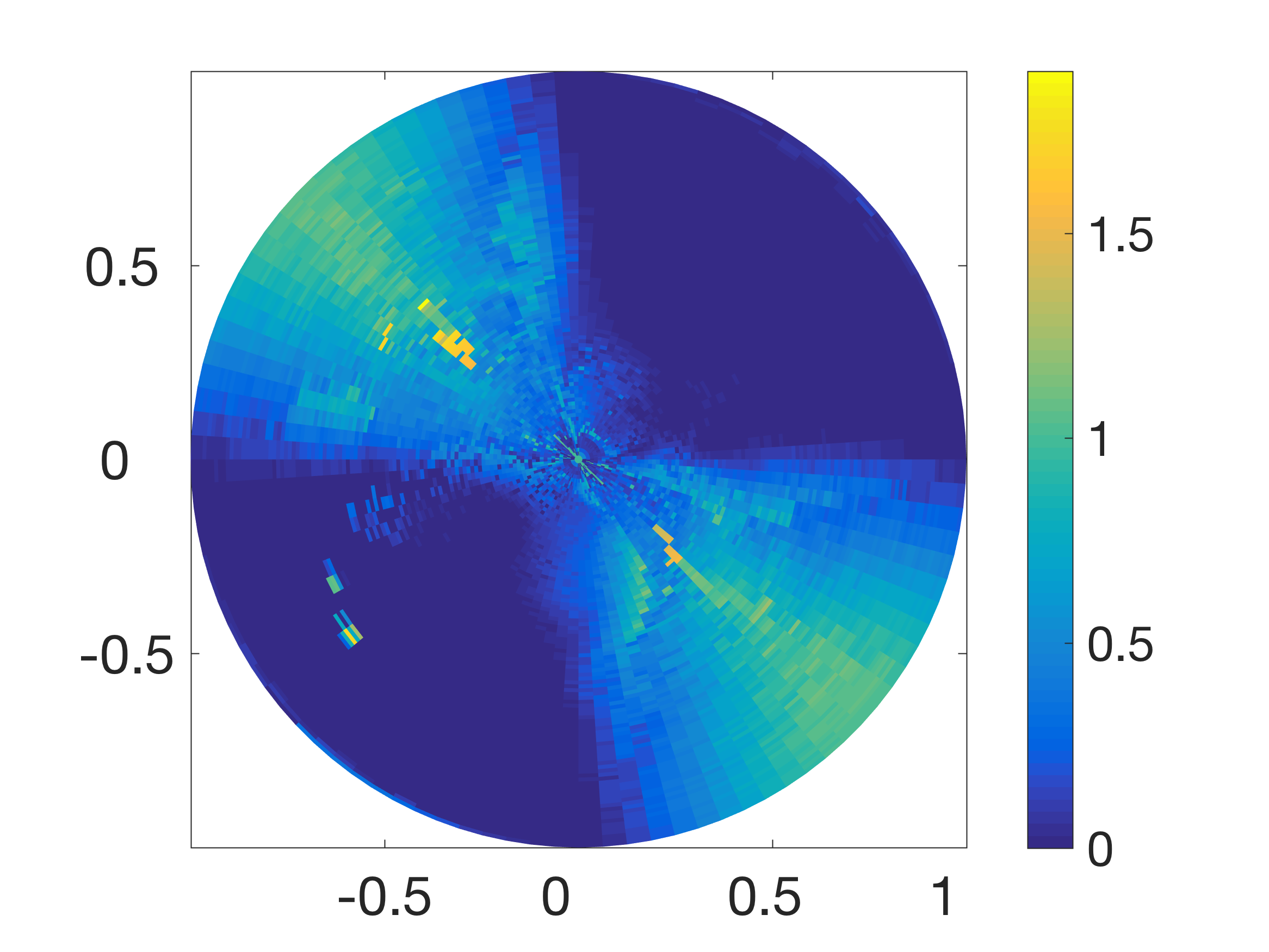}} \\
  \subfloat[][]{\includegraphics[width=6cm]{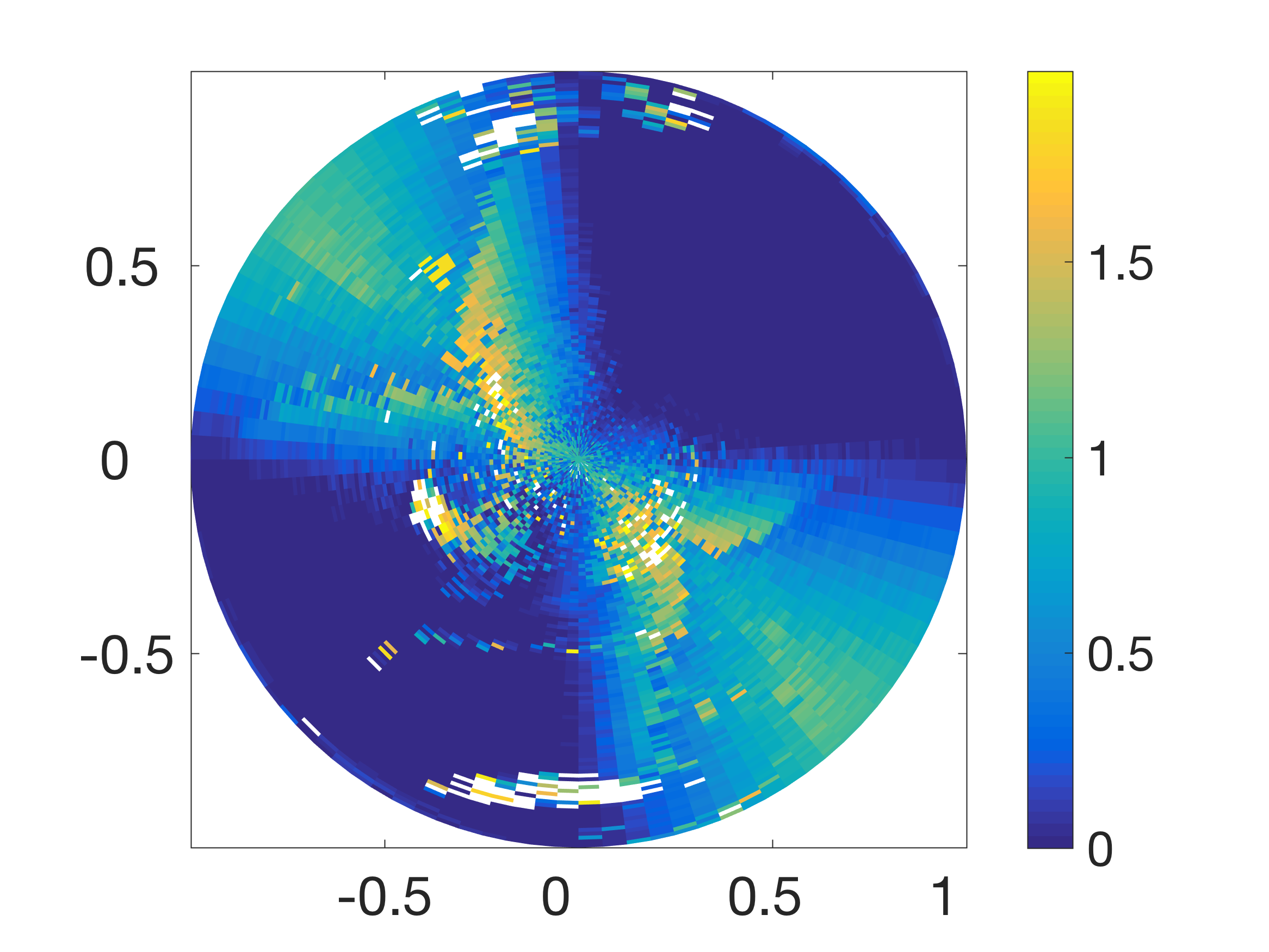}} \quad
  \subfloat[][]{\includegraphics[width=6cm]{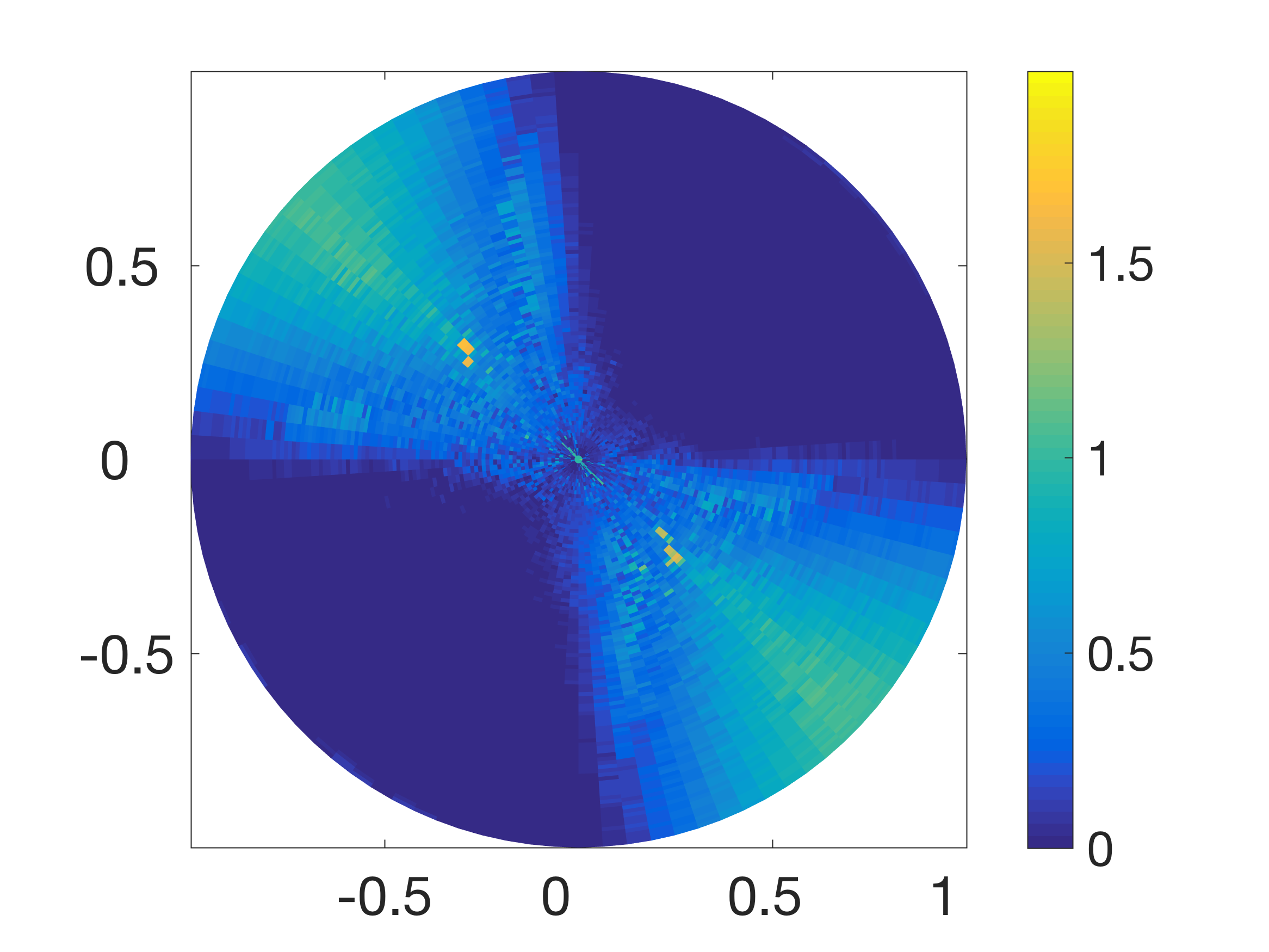}}
  \caption{The minimal quantum relative entropy for different initial
    conditions. (a) $\mathbf{a} =0$. (b) $\mathbf{a}=1$. (c)
    $\mathbf{a}=2$, and the white color denote the calculated minimal
    quantum relative entropy is larger than $2$ at the point. (d) The
    minimum quantum relative entropy taken over the cases (a), (b),
    and (c). }
  \label{fig:qbm3q1h}
\end{figure}

We still study the target states $|\psi(r,\phi)\rangle$ in Eq.~\eqref{eq:83}. The
numerical results from the BFGS algorithm are shown in
Fig.~\ref{fig:qbm3q1h}, where the results for different initial
parameters $\mathbf{a}=0$, $\mathbf{a}=1$, and $\mathbf{a}=2$ are
shown in Fig.~\ref{fig:qbm3q1h}~(a), (b), and (c) respectively.
Obviously, when the hidden subsystem, the third qubit, is involved,
the relative entropy will have many local minimums, rather than only
one minimum in the case without hidden subsystems. The minimal quantum
relative entropy taken over the cases (a), (b), (c) is shown in
Fig.~\ref{fig:qbm3q1h}~(d). Compared the results in
Fig.~\ref{fig:qbm3q1h} (d) and Fig.~\ref{fig:qbm2q0h}, we observe that
the relative entropy decreases obviously, e.g. the maximal minimum
quantum relative entropy at $\{r=\sqrt{2}/2,\phi=\frac{3\pi}{4}\}$
decreases from $2$ to about $1.00$.

The numerical results in Fig.~\ref{fig:qbm3q1h} (d) and
Fig.~\ref{fig:qbm2q0h} show that there are two symmetric axes: $y=x$
and $y=-x$ in the minimal quantum relative entropy of
$|\psi(r,\phi)\rangle$. Now let us analyze the symmetry in the quantum Boltzmann
machine for the target states $\sigma_{\ast}(\mathbf{r})$.

First, we notice that the local unitary transformation $U_{v}$
simultaneously acting on $\sigma_{\ast}(\mathbf{r})$ and
$\sigma(\mathbf{a})$ does not change the quantum relative entropy, i.e.,
\begin{equation}
  \label{eq:25}
  S(U_{v}\sigma_{\ast}(\mathbf{r})U_{v}^{\dagger}|U_{v}\sigma(\mathbf{a})U_{v}^{\dagger})
  = S(\sigma_{\ast}(\mathbf{r})|\sigma(\mathbf{a})).
\end{equation}
In addition, a local unitary transformation $U_{h}$ acting on the
hidden subsystem does not change the reduced state
$\sigma(\mathbf{a})$. In other words, the quantum relative entropy
$S(\sigma_{\ast}(\mathbf{r})|\sigma(\mathbf{a}))$ is invariant under arbitrary
unitary transformation $U_{v}\otimes U_{h}$.

To make the unitary transformation $U_{v}\otimes U_{h}$ become a symmetric
operation, we require that
\begin{align}
  \label{eq:26}
  &  U_{v} \sigma_{\ast}(\mathbf{r}) U_{v}^{\dagger} =
    \sigma_{\ast}(\mathbf{r}^{\prime}), \\
  & U_{v}\otimes U_{h} H(\mathbf{a}) U_{h}^{\dagger}\otimes U_{v}^{\dagger} = H(\mathbf{a}^{\prime}),
\end{align}
where $H(\mathbf{a})$ is the Hamiltonian in Eq.~\eqref{eq:5}. In other
words, we require that the unitary transformation keeps the forms of
$\sigma_{\ast}(\mathbf{r})$ and $H(\mathbf{a})$, and it only changes the
parameters from \{$\mathbf{r}, \mathbf{a}$\} to
\{$\mathbf{r}^{\prime}$, $\mathbf{a}^{\prime}$\}.

As an example, we will show that $U_{v}=X_{1}\otimes X_{2}$ and
$U_{h}=I_{3}$ is a symmetric operation. Note that
\begin{equation}
  \label{eq:114}
  |\psi(r,\frac{\pi}{2} - \phi)\rangle = X_{1} \otimes X_{2} |\psi(r,\phi)\rangle,
\end{equation}
and
\begin{align}
  \label{eq:116}
  H(\mathbf{a}^{'}) & = X_{1} \otimes X_{2} H(\mathbf{a}) X_{1} \otimes X_{2} \\
                    & = - \sum_{i=1}^{3} a_{i} X_{i} + \sum_{i=1}^{2} a_{i+3} Z_{i} - a_{6} Z_{3} -
                      a_{9} Z_{1} \otimes Z_{2} + a_{7} Z_{2} \otimes Z_{3} + a_{8} Z_{1} \otimes Z_{3},
\end{align}
which gives the relation between $\mathbf{a}^{'}$ and $\mathbf{a}$.
Thus we obtain
\begin{equation}
  \label{eq:117}
  S(\sigma_{\ast}(r,\frac{\pi}{2}-\phi)|\sigma(\mathbf{a}^{'})) = S(\sigma_{\ast}(r,\phi)|\sigma(\mathbf{a})).
\end{equation}
Because the above relation is valid for arbitrary $\phi$ and
$\mathbf{a}$, we conclude that
\begin{equation}
  \label{eq:118}
  \min_{\mathbf{a}^{'}}
  S(\sigma_{\ast}(r,\frac{\pi}{2}-\phi)|\sigma(\mathbf{a}^{'})) =
  \min_{\mathbf{a}} S(\sigma_{\ast}(r,\phi)|\sigma(\mathbf{a})),
\end{equation}
i.e.,
\begin{equation}
  \label{eq:27}
  S_{m}(r,\frac{\pi}{2}-\phi) = S_{m}(r,\phi).
\end{equation}

Similarly, for the symmetric operation $U_{v}=Y_{1}\otimes Y_{2}$ and
$U_{h}=I_{3}$, we can show that
\begin{equation}
  \label{eq:119}
  S_{m}(r,\frac{3\pi}{2}-\phi) = S_{m}(r,\phi).
\end{equation}
This implies that the states $(r,\phi)$, $(r,\frac{\pi}{2}-\phi)$,
$(r,\frac{3\pi}{2}-\phi)$ and $(r,\pi+\phi)$ has the same minimum quantum
relative entropy. Because of the symmetry, we only need to consider
the states for $\phi\in[\frac{\pi}{4},\frac{3\pi}{4}]$.

In fact, up to a phase factor, all the symmetry operations can be
written as
\begin{equation}
  \label{eq:120}
  \{X_{1}\otimes X_{2},Y_{1}\otimes Y_{2},Z_{1}\otimes Z_{2},I_{1}\otimes
  I_{2}\} \otimes \{X_{3}, Y_{3}, Z_{3}, I_{3}\}.
\end{equation}

Let us consider the symmetric operation $U_{v}=I_{1}\otimes I_{2}$ and
$U_{h}=X_{3}$. Then we have
\begin{equation}
  \label{eq:13}
  S(\sigma_{\ast}(\mathbf{r})|\sigma(\mathbf{a}^{\prime})) =
  S(\sigma_{\ast}(\mathbf{r})|\sigma(\mathbf{a}))
\end{equation}
with
\begin{equation}
  \label{eq:14}
  H(\mathbf{a}^{\prime}) = - \sum_{i=1}^{3} a_{i} X_{i} -
  \sum_{i=4}^{5} a_{i} Z_{i} - a_{9} Z_{1}\otimes Z_{2}
  + a_{6} Z_{3} + a_{8} Z_{1} \otimes Z_{3} + a_{7} Z_{2} \otimes Z_{3}.
\end{equation}
This implies that if $\mathbf{a}$ corresponds to a minimum, then
$\mathbf{a}^{\prime}$ must also be a minimum. If the hidden subsystem
decrease $S_{m}(\mathbf{r})$, then $d_{13}d_{23}\neq 0$, which leads to
$\mathbf{a}^{\prime}\neq \mathbf{a}$, i.e., there are at least $2$
degenerate parameter vectors $\mathbf{a}$ for the same state.

\begin{figure}[ht]
   \centering
   \subfloat[][]{\includegraphics[width=6cm]{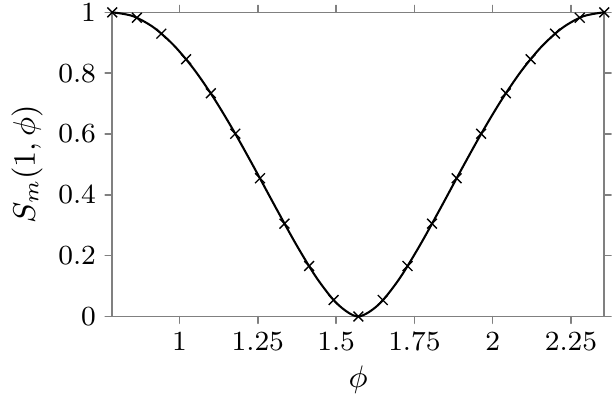}} \quad
   \subfloat[][]{\includegraphics[width=6cm]{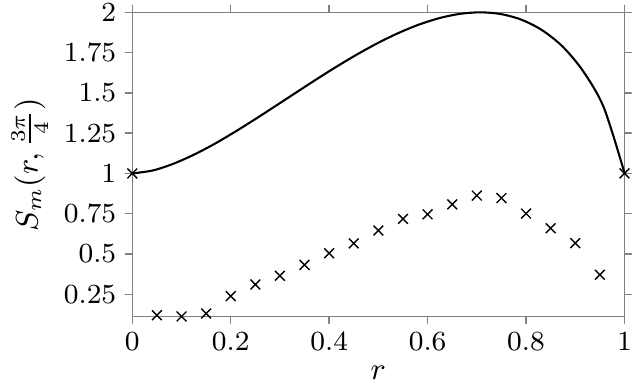}}

   \caption{The minimal quantum relative entropy in the two cases: (a)
     $r=1$; (b) $\phi=\frac{3\pi}{4}$. The analytical results without the
     hidden subsystem are plotted as the solid lines, and the
     numerical results with the hidden subsystem are plotted as cross
     marks. \label{fig:ex2h}}
 \end{figure}

In general, when there is a hidden subsystem, there are many local
minimums, which makes the numerical results from our BFGS algorithm
strongly depends on the initial parameter vector $\mathbf{a}_{0}$, and
it is hard to obtain the global minimum. Here for each target state,
we take the minimum from the numerical results with $1000$ random
initial parameter vectors $\mathbf{a}_{0}$. The numerical results are
shown in Fig.~\ref{fig:ex2h}. For the states $|\psi(1,\phi)\rangle$, the hidden
subsystem does not increase the power of the quantum Boltzmann
machine. However, the hidden subsystem significantly increases the
power of the quantum Boltzmann machine for the states
$|\psi(r,\frac{3\pi}{4})\rangle$ with $0<r<1$.

 \section{Discussion and summary}
 \label{sec:summary}

 We developed the optimization BFGS algorithm for quantum Boltzmann
 machine based on analytical results on the first order derivatives of
 quantum relative entropy. Then we apply the algorithm to the
 two-qubit target states, a family of states with two parameters with
 the transverse Ising model without or with hidden layers.

 For the machine learning model without hidden layers, the numerical
 results show that the target states in the first and the third
 quadrants are perfectly represented by the Boltzmann machine, while
 the target states in the second and the fourth quadrants are not. We
 use the numerical range of related observables for any states forms a
 convex body, and we further show that the target states in the first
 and third quadrants lie in the surface of the convex body, but those
 states in the second and the fourth quadrants lie in the interior of
 the convex body. Since a Hamiltonian specifies a hyperplane in the
 case, the states in the surface can be regarded as the ground states
 of the Hamiltonian, which can be used to construct the Boltzman
 machine. Thus the geometric picture gives a clear physical
 explanation of the quadrant-dependent behavior of the minimum quantum
 relative entropy.

 For the quantum Boltzmann machine with or without hidden layers,
 there are two symmetric axes: $y=x$ and $y=-x$ in the minimal quantum
 relative entropy of $|\psi(r,\phi)\rangle$. We clarify how to define the
 symmetric operation from the invariance of quantum relative entropy
 under local unitary transformations. Then we prove the symmetry in
 our problem by explicitly constructing the symmetric operations
 corresponding to these two symmetric axes. Furthermore, we also use a
 symmetry argument to show that there are many local minimums for the
 Boltzmann machine with hidden layers.

 In summary, we study a family of target states with the quantum
 Boltzmann machine by developing the BFGS algorithm, and explain the
 numerical results through the geometric method and the symmetry
 analysis. We hope that our results can lead to the global
 characterization on the target states, and increases the
 understanding of quantum machine
 learning~\cite{Nature549,Schuld2017,2014arXiv1412.3489W,2016arXiv161205695C}.

\begin{acknowledgments}
  This work is supported by NSF of China (Grant Nos. 11475254 and
  11775300), NKBRSF of China (Grant No. 2014CB921202), the National
  Key Research and Development Program of China (2016YFA0300603).
\end{acknowledgments}

\bibliographystyle{unsrt} \bibliography{qbm}

\end{document}